\documentclass[twocolumn]{aastex631}

\usepackage{natbib}

\usepackage{graphicx}
\usepackage[space]{grffile}
\usepackage{latexsym}
\usepackage{amsfonts,amsmath,amssymb}
\usepackage{url}
\usepackage[utf8]{inputenc}
\usepackage{fancyref}
\usepackage{hyperref}
\hypersetup{colorlinks=true,citecolor=blue,urlcolor=blue,pdfborder={0 0 0},}
\usepackage{textcomp}
\usepackage{multirow,booktabs}
\usepackage{subcaption}
\usepackage{xspace}
\newcommand{\mr}{\mathrm}
\shorttitle{Exoplanet Oblateness}
\shortauthors{Berardo \& de Wit 2022}
\begin{document}

\title{On The Effects of Planetary Oblateness on Exoplanet Studies}

\author{David Berardo}
\altaffiliation{Department of Physics and Kavli Institute for Astrophysics and Space Research, Massachusetts Institute of Technology, Cambridge, MA 02139, USA}
\altaffiliation{FRQNT Doctoral Research Scholarship}

\author{Julien de Wit}
\altaffiliation{Department of Earth, Atmospheric and Planetary Sciences, Massachusetts Institute of Technology, Cambridge, MA 02139, USA}

\correspondingauthor{David Berardo}
\email{berardo@mit.edu}

\begin{abstract}
	    
	When studying transiting exoplanets it is common to assume a spherical planet shape. However short rotational periods can cause a planet to bulge at its equator, as is the case with Saturn whose equatorial radius is almost 10\% larger than its polar radius. As a new generation of instruments comes online, it is important to continually assess the underlying assumptions of models to ensure robust and accurate inferences. We analyze bulk samples of known transiting planets and calculate their expected signal strength if they were to be oblate. We find that for noise levels below 100ppm, as many as 100 planets could have detectable oblateness. We also investigate the effects of fitting spherical planet models to synthetic oblate lightcurves. We find that this biases the retrieved parameters by several standard deviations for oblateness values $>$ 0.1-0.2. When attempting to fit an oblateness model to both spherical and oblate lightcurves, we find that the sensitivity of such fits is correlated with both the SNR as well as the time sampling of the data, which can mask the oblateness signal. For typical values of these quantities for Kepler observations, it is difficult to rule out oblateness values less than $\sim$0.25. This results in an accuracy wall of 10-15$\%$ for the density of planets which may be oblate. Finally, we find that a precessing oblate planet has the ability to mimic the signature of a long-period companion via transit timing variations, inducing offsets at the level of 10s of seconds.

	%results: given the bulk sample of exoplanets, many could have potentially been observable with K2, and many should be with jwst. For those that weren't, it's possible that they were missed because a circular model does just fine at fitting them. When you fit for eccentricity, this behaviour seems to be diminished. Regarding density, its possible a small # of planets have their density wrong by several sigma. Known exoplanets could mimic ttvs on the order of 10's of seconds, although on very long timescales. When looking at synthetic populations of spherical / oblate planets, clear trends emerge when trying to detect population oblateness.
\end{abstract}


\section{Introduction}
\label{sec:intro}
%The goal of this paper is to study the feasibility of detecting oblate planets. We look at correlations with other observable, as well as the possibility of finding oblate planets in surveys such as K2 and TESS which have yet to be mined for such planets.

Young planets which have yet to come into tidal equilibrium with their host star may be rotating at speeds sufficient to cause their equators to bulge and thus deviate substantially from a sphere.  When analyzing the lightcurves of transiting exoplanets it is typical to assume that a planet is perfectly spherical, as is the case of the many fitting routines which implement the transit models of \cite{mandel:2002}. Within our own solar system however, we observe that the equatorial radii of Saturn and Jupiter are $9.8\%$ and $6.5\%$ larger than their polar radii. Planets can become distorted due to tidal forces from their host star or, as is the case with Jupiter and Saturn, due to rapid rotation, causing them to bulge uniformly about their rotation axis. The amount by which a planet is oblate is tied to its rotation rate as well as its internal structure, which determines its deformability. The ability to measure the oblateness of a planet would thus allow us to better understand its current rotational properties and internal dynamics, as well as its formation and evolution (which determine its current rotation) \citep{lissauer:1995}.
\begin{figure*}[!ht]
	\centering
	\includegraphics[scale=0.28]{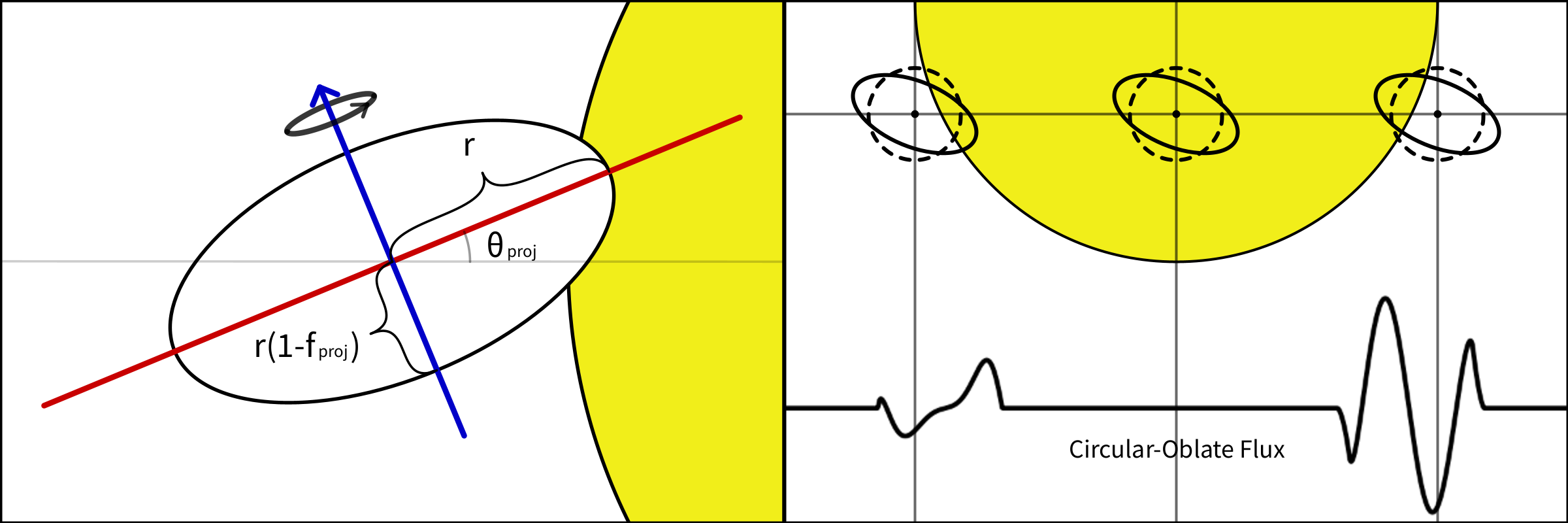}
	\caption{Left: A schematic showing the configuration of the planet and all relevant quantities. The blue (polar) minor axis is the planet's axis of rotation, which is scaled down by a factor of $(1-f_{\mr{proj}})$ relative to the red (equatorial) major axis. The planet may also be tilted by an angle $\theta_{\mr{proj}}$ relative to its orbital plane. Note these quantities are the sky-plane projected values an observer would see. Right: The difference in the lightcurve of an oblate planet (solid) to a spherical planet (dashed) with the same projected area. For a uniformly bright star, there is no difference once the planets are fully inside the stellar disk. Note the asymmetry induced by the orbit having a non-zero impact parameter.}
	\label{fig:oblateness diagram}
	
\end{figure*}
The effect of oblateness on the lightcurve of a transiting planet has previously described in the works of \cite{seager:2002}, \cite{barnes:2003},  \cite{carter:2010a}, and \cite{carter:2010b}. These works highlight the features of an oblate lightcurve, which is primarily a variation in the ingress and egress of their transits on the order of 10's to 100's of ppm for the most optimal of planets. These properties outline the difficulties in detecting oblateness through lightcurve variations, which has thus far proven to be a difficult task. In \cite{zhu:2014}, short cadence planets in the \textit{Kepler} survey were searched, yielding a tentative result of oblateness for the planet Kepler-39b (KOI432.01). A further search was conducted in \cite{biersteker:2017}, which attempted to detect oblateness using depth variations across multiple transits as an oblate planet precesses over time. This yielded yet another tentative result of oblateness for the warm Saturn Kepler-427b. Recently \cite{akinsanmi:2020} studied the potential of spectroscopic measurements to detect oblateness, finding that a combination of photometry and spectroscopy has the potential to provide more accurate and precise oblateness measurements.

In section \ref{sec:model} we outline the relevant quantities and processes that control planetary oblateness. We then analyze a bulk sample of planets to determine how many should have detectable oblateness for varying levels of sensitivity. In section \ref{sec:biases} we fit spherical planet models to synthetic oblate lightcurves to study biases on the retrieved transit parameters. As shown in \citet{dewit:2012}, orbital parameters can compensate for shape and brightness distribution effects on the shape of transit and eclipses, an effect which we seek to further understand. In section \ref{sec:fitting}, we fit oblate models to synthetic and real planets in order to determine the regimes in which such models can properly measure or rule out oblateness, as well as to determine if populations can be separated into oblate/non-oblate based on properties such as orbital period. Finally in section \ref{sec:disc} we discuss the effects oblateness may have on the precision of measured planet densities, as well as the ability for time-varying oblateness to mimic transit timing variations due to precession of a planet's orbital axis.

\section{Description and Physical Background of Oblateness}
\label{sec:model}

Oblateness refers to the amount by which the equatorial radius of a planet differs from its polar radius. A common cause of planet shape distortions are tidal forces which induce a bulge towards the host star \citep{love:1911}. In our case we are interested in oblateness induced by a planets rotation, which causes its equator to bulge out perpendicular to its rotation axis. Standard notation uses the parameter $f$ for oblateness, which is defined as:

\begin{equation}
	f = \frac{R_{eq} - R_{pol}}{R_{eq}}
\end{equation}

where $R_{eq}$ and $R_{pol}$ are the equatorial and polar radii of a planet \citep{murray:1999}. Thus oblateness ranges from 0 (for a perfectly spherical planet) to 1 (for a planet completely flattened along the radial direction). In principle $f$ could also be negative for the case of a prolate planet, whose polar radius is larger than its equatorial radius. While we restrict the oblateness to be a positive value in this work, we do allow its rotational axis to be tilted up to 90 degrees relative to its orbital axis, which would cause an oblate planet to appear prolate.  For the planets in our own solar system, $f$ ranges from near zero for the rocky planets such as mercury and earth, to approximately 0.065 and 0.1 for Jupiter ($P_{Rot} = 9.93\ \mr{hrs}$) and Saturn ($P_{Rot} = 10.65\ \mr{hrs}$) respectively\footnote{Values taken from \url{https://nssdc.gsfc.nasa.gov/planetary/planetfact.html}}. The rotational period of a planet and its oblateness are related by \citep{hubbard:1984}

\begin{equation}
	\label{eq:prot}
	P_{rot} = 2\pi \sqrt{\frac{R_{eq}^3}{GM_p\left(2f-3J_2\right)}}
\end{equation}

where $M_p$ is the planet's mass, and $J_2$ is its quadrupole moment, which is around 0.015 for Jupiter and Saturn, around 0.001 for Earth and Mars, and $<$ 0.00001 for Mercury and Venus.

The planet may also be tilted relative to its orbital plane by two obliquity angles $\theta$ and $\phi$. When an ellipsoid is rotated in three dimensions and projected onto the sky plane we observe an ellipse which we can characterise by a \textit{projected} oblateness $f_{\mr{proj}}$and a \textit{projected} obliquity $\theta_{\mr{proj}}$, as shown in the left panel of figure \ref{fig:oblateness diagram}. The projected values are related to the true values by

\begin{subequations}
	\label{eq:proj_eqs}
	\begin{equation}
		\label{eq:f_proj}
		f_{\mr{proj}} = 1 - \sqrt{\mr{sin}^2\theta' + (1-f)^2\mr{cos}^2\theta'}
	\end{equation}
	\begin{equation}
		\label{eq:theta_proj}
		\mr{tan}\theta_{\mr{proj}} = \mr{tan}\theta\mr{sin}\phi
	\end{equation}
	\begin{equation}
		\label{eq:theta_prime}
		\mr{cos}^2\theta' = \mr{sin}^2\theta\mr{sin}^2\phi + \mr{cos}^2\theta    
	\end{equation}
\end{subequations}

\begin{figure*}
	\centering
	\begin{subfigure}[t]{0.48\textwidth}
		\centering
		\includegraphics[width=\textwidth]{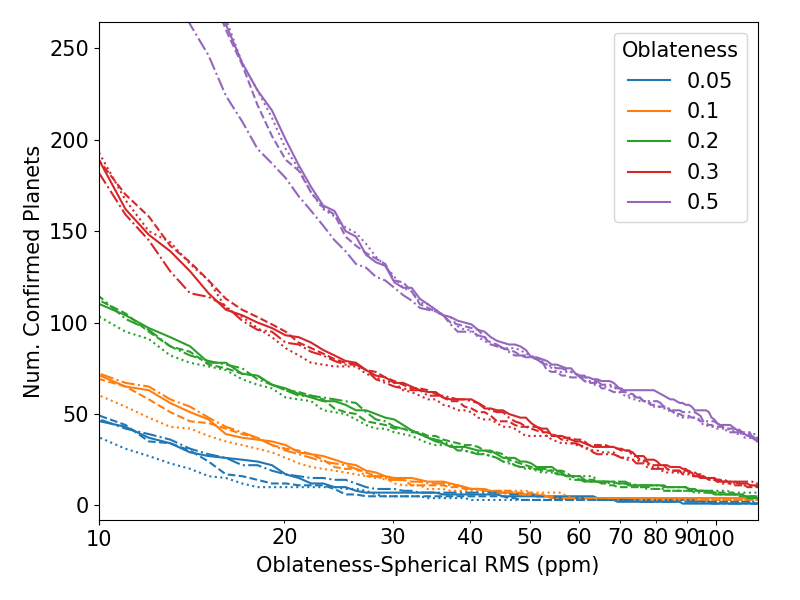}
		\caption{The number of confirmed transiting planets whose lightcurve would vary by a given amount if they had a certain level of oblateness, with a one minute integration time. The x-axis represents the Root-Mean-Square (RMS) deviations during ingress. The solid, dashed, dotted, and dash-dot lines represent obliquities of 0, 45, 60, and 90 degrees respectively.}
		\label{fig:known planet oblateness}
	\end{subfigure}
	\hfill
	\begin{subfigure}[t]{0.48\textwidth}
		\centering
		\includegraphics[width=\textwidth]{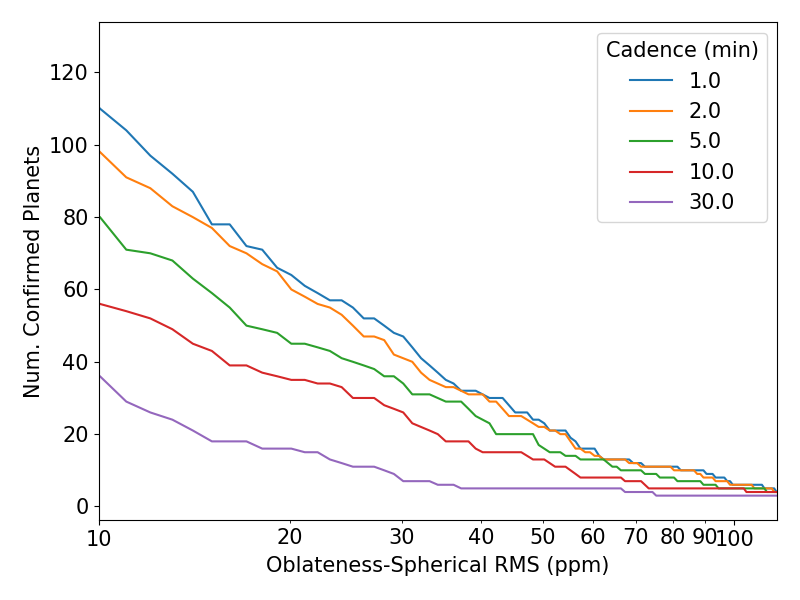}
		\caption{For the same sample of planets as in a), we keep the oblateness fixed to 0.2 in all cases. The curves correspond to different observing cadences which smooth out the oblateness signal.}
		\label{fig:cadence effect on ob amp}
	\end{subfigure}
	\caption{The expected oblateness of confirmed transiting planets}
	\label{fig:three graphs}
\end{figure*}
This implies that for most planets what will be observed is an oblateness value that is smaller than their true oblateness, diminishing the deviations from a spherical planet. If one assumes that the orbital axis is randomly oriented (i.e. uniformly distributed on the unit sphere), then the median projected oblateness would be between 75\% to 70\% of the true oblateness (see appendix A). Constraints on the amount by which a planet's rotation can be tilted away from the normal vector of its orbital plane would allow us to put limits on the effect of projection. However in our own solar system we observe Uranus to have its rotational axis oriented by $97.77^{\circ}$ relative to its orbital plane, providing us with the two extremes of perfect alignment (as is the case for Mercury) as well as perfect misalignment, along with several intermediate values.

The observable effect of an oblate planet is illustrated in the right panel of figure \ref{fig:oblateness diagram}. During ingress and egress of a transit, two planets with differing shapes will block differing amounts of light from their host star. More specifically, the time-dependence of the amount of light they block will vary, which affects the shape of the lightcurve as the planet enters and exists the limb of the stellar disk. Once the planet is fully inside the disk of the star, the dominating factor in the amount of light blocked is the total projected area of the planet. Stellar heterogeneities such as limb darkening  and star spots would still produce differing signals for spherical and non-spherical planets, however the effect is much weaker when compared to the differences during ingress and egress, which may be be upwards of several 100's of parts per million (ppm). It is thus critical when searching for oblateness to ensure that the limbs of the transit are sufficiently sampled in time.

\subsection{The observability of oblateness for known planets}
\label{sec:known planet ob level}

Currently, the detection of oblateness is limited to a handful of tentative measurements among the most suitable planet candidates. Previous studies such as \cite{zhu:2014} and \cite{biersteker:2017} have used the \textit{Kepler} survey as their pool of observations, which have provided us with some of the highest quality, as well as longest baseline, transit observations to date.

The goal of this section is to quantify the average strength of an oblateness signal that one expects to see across all known planets. This allows us to determine both the number of expected planets whose oblateness may be measured with upcoming instruments, as well as the potential to place upper limits on the oblateness of planets if no such signal is seen at increased signal to noise ratios (SNR). The amplitude of an oblateness signal depends on properties such as the planet-to-star area ratio, inclination, impact parameter etc. in non-linear ways, and so planets must be assessed on an individual basis. We compiled a list of all known transiting planets, taken from the exoplanet archive table of confirmed planets \citep{confirmedplanets}. As will be discussed in section \ref{sec:metric}, it is expected that planets with short orbital periods ($\lesssim$ 15 days) will be rotating too slowly to be oblate. We thus cut out planets with a period of less than 10 days, in order to exclude planets which we confidently do not expect to exhibit oblateness so as to not inflate the number of potential candidates. We also exclude planets which do not have reported values for any of the transit parameters necessary to calculate a lightcurve model.

This leaves us with a sample of $\sim$900 planets. For each of these we then generate synthetic lightcurves for a spherical planet (using the Batman package of \cite{kreidberg:2015b}), as well for varying levels of oblateness. The oblateness is kept to be less than 0.5 ($r_{eq} \le 2 r_{pol}$) for stability considerations. Equating the centripetal and gravitational forces at the planets surface

\begin{equation}
	\frac{GM}{r_{eq}^2} = \frac{v^2}{r}
\end{equation}

and substituting equation \ref{eq:prot} into $v = 2\pi r_{eq} / P_{rot}$ to relate the velocity at the surface to the oblateness parameter f, we find

\begin{equation}
	f_{crit} = (1 + 3J_2)/2
\end{equation}

which is  approximately 0.5, for $3J_2 << 1$. To obtain values of f larger than 0.5, the necessary rotational velocity of the planet would overcome its gravitational attraction and become unstable.

We generate 1-minute cadence lightcurves of oblate planets using a Monte-Carlo integration routine, based off appendix A of \cite{carter:2010a}, which includes a quadratic limb darkening law\footnote{Using the corrected versions of equations (B5) and (B6) noted in \cite{zhu:2014}, $r=\sqrt{u + (1-u)(a_1/a_2)^2}$ and $\theta = (1-v)\theta_1+ v\theta_2$}. The uncertainty in the model itself is at the level of 1-2 ppm (set by the number of points used in the integration). 

When generating oblate lightcurves of a given planet, we fix the transit parameters to their measured values. The only parameter we vary is the equatorial radius of the planet, which we adjust so that the depth of the transit remains constant. The transit depth of a planet of oblateness f is given by

\begin{equation}
	\delta_{ob} = r_{eq}r_{polar} = r_{eq}r_{eq}(1-f)
\end{equation}

and so if we were to keep the radius the same, the depth between an oblate and spherical planet would differ in proportion to f. Given that during the center of transit there is little to no information regarding the shape of the planet (depending on the degree of limb darkening) we set $\delta_{ob} = \delta_{circ}$, which gives the relation between the equatorial radius of an oblate and spherical planet as

\begin{equation}
	r_{ob} = r_{circ} / \sqrt{1-f}
\end{equation}

We calculate the difference between the spherical and oblate lightcurves, and then calculate the root mean square of the difference during ingress and egress, which we show in figure \ref{fig:known planet oblateness}. Alternatively, we could have chosen the amplitude or maximum difference between the lightcurves. However for many planets which have low time-sampling during ingress / egress, it would be misleading to report what is usually a single high point as the overall strength of the oblateness signal.

We find that between 10ppm - 100ppm is when most planets begin to exhibit oblateness. For nearly maximal value of f$\sim$0.5 we find as many as 50 planets with RMS deviations above 100ppm. However this is an extreme case, and as mentioned due to projection effects it is unlikely many planets would display such a high level of oblateness. At a signal amplitude of approximately 50ppm, there are on the order of 30 planets which would exhibit variations at this level if they were to have an oblateness factor of 0.2, and almost double this number of planets if they had an oblateness of 0.3. Going down to 10ppm, we see these numbers roughly double again. %At this level, it would take only a small percentage of planets to be oblate for us to be able to make a definitive measurement of extra-solar oblateness. 

In addition to oblateness, we also vary the obliquity angle of the planets. We find that when considering a bulk population the obliquity does not significantly change the number of observable planets at a given signal level. As shown in the right panel of figure \ref{fig:oblateness diagram}, it is possible for the signal to vary significantly between ingress and egress, indicating that transit asymmetry may be a potential sign of oblateness. If we calculate the RMS separately for ingress and egress and compare them, we find that the ratio of their values is on average between 0.8 and 0.9, depending on the oblateness value. For planets with sufficiently sampled limbs this effect would be an additional marker of oblateness.

We also study the effect of observing cadence (i.e. the integration time between data points), which is shown in figure \ref{fig:cadence effect on ob amp}. Given that oblateness primarily causes variations during ingress and egress, the number of observations during these parts of the transit directly impacts the ability to detect oblateness. We hold the oblateness level fixed at 0.2 and alter the duration of the observing cadence, ranging from 1 minute to 30 minutes, as a reflection of the Kepler observing cadence modes. We see in this case that for a given signal level, the number of planets which exhibit oblateness decreases by more than half as the integration time increases. We note that for most of the planets in our sample, a one minute observing cadence corresponds to about 5 - 50 data points during the transit limbs, with a median of 16 data points. The effect of time sampling, which smooths and distorts the shape of a transit, is often not considered when discussing oblateness. The significance of this is analysed further in section \ref{sec:sensitivity}.

\subsection{Where are all the oblate planets?}

We have shown that as many as several tens of exoplanets should have levels of oblateness in the 30-100ppm range at observing cadences of $<$ 10 minutes. While a small signal, this is certainly within the capabilities of facilities such as \textit{Kepler}, as shown in figure 5 of \cite{koch:2010}. To date, there is only a small handful of planets which have had tentative detections of oblateness, as mentioned in section \ref{sec:intro}. This begs the question of why there haven't been a larger number of conclusive oblateness measurements. One explanation for this is that exoplanets are simply not oblate, which would make the case of Saturn's oblateness an extreme outlier, or that oblate planets happen to be found around dimmer stars with lower photometric precision. Another option which we will explore in the next section is that the transit parameters of associated with a spherical planet model are able to mimic the signature of oblateness and thus compensate for it.

\section{Mapping the Degeneracies Between Oblateness and Transit Parameters}
\label{sec:biases}

As shown in the previous section, the typical deviations induced by oblateness will be in the best case a few hundred ppm, and much more typically well below 100 ppm. The additional fact of the signal being confined solely to the limbs of a transit means that for many planets, depending on the duration of the transit and the cadence at which the observations are taken, only a handful of data points may be useful in detecting oblateness. It is thus not surprising that when analysing an exoplanet transit the standard practice is to assume a perfectly spherical planet. Nevertheless, we propose that even in cases of non-detection oblateness may have a significant effect on parameter retrieval.

\begin{figure*}
	\centering
	\includegraphics[scale=0.65]{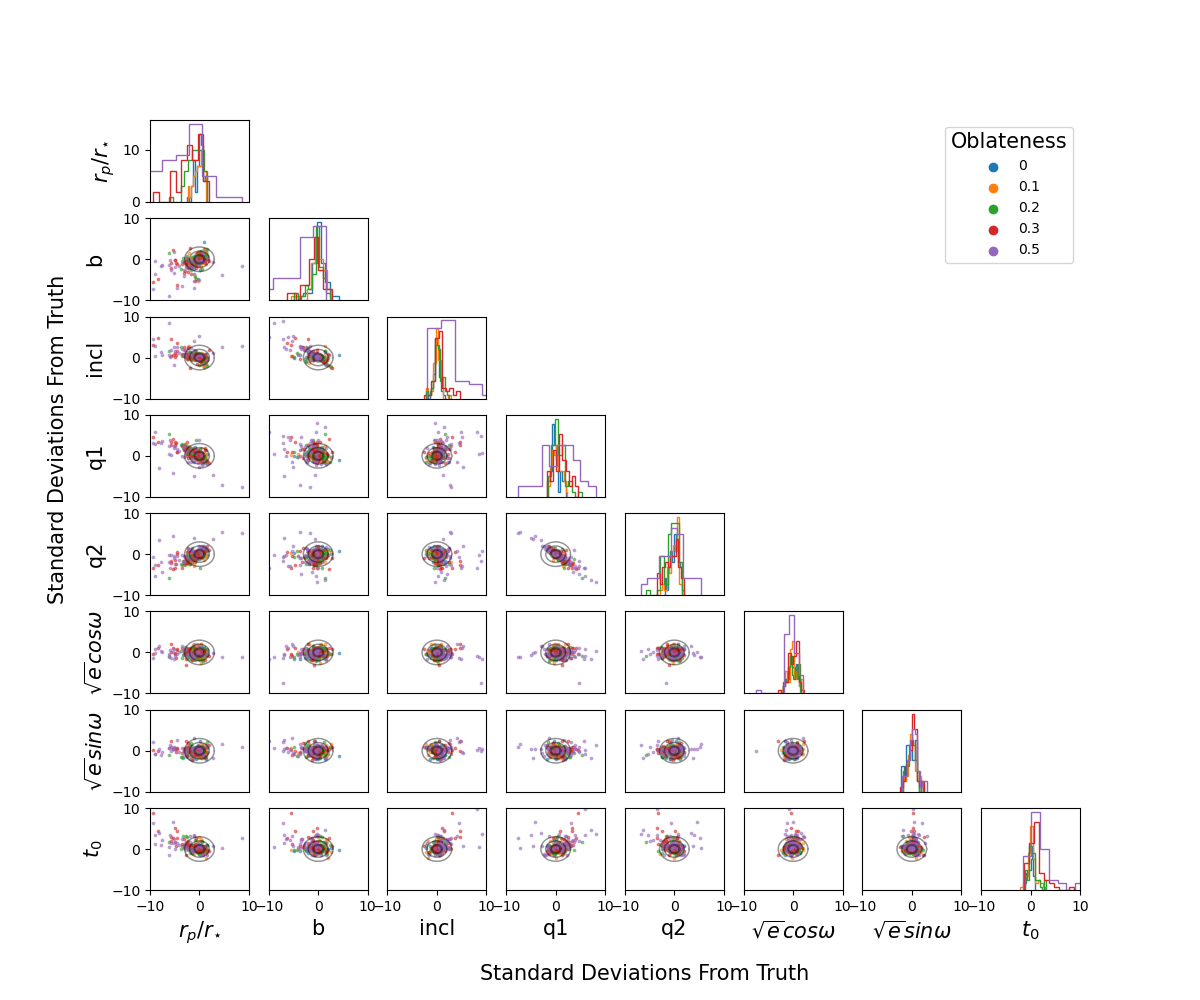}
	\caption{Corner plot showing the deviations for each parameter from their truth values, in units of standard deviations of the posterior. Colors represent different levels of oblateness. The black circles show boundaries of 1, 2, and 3 standard deviations. }
	\label{fig:corner plots}
\end{figure*}
\begin{figure*}[ht]
	\centering
	\includegraphics[scale=0.75]{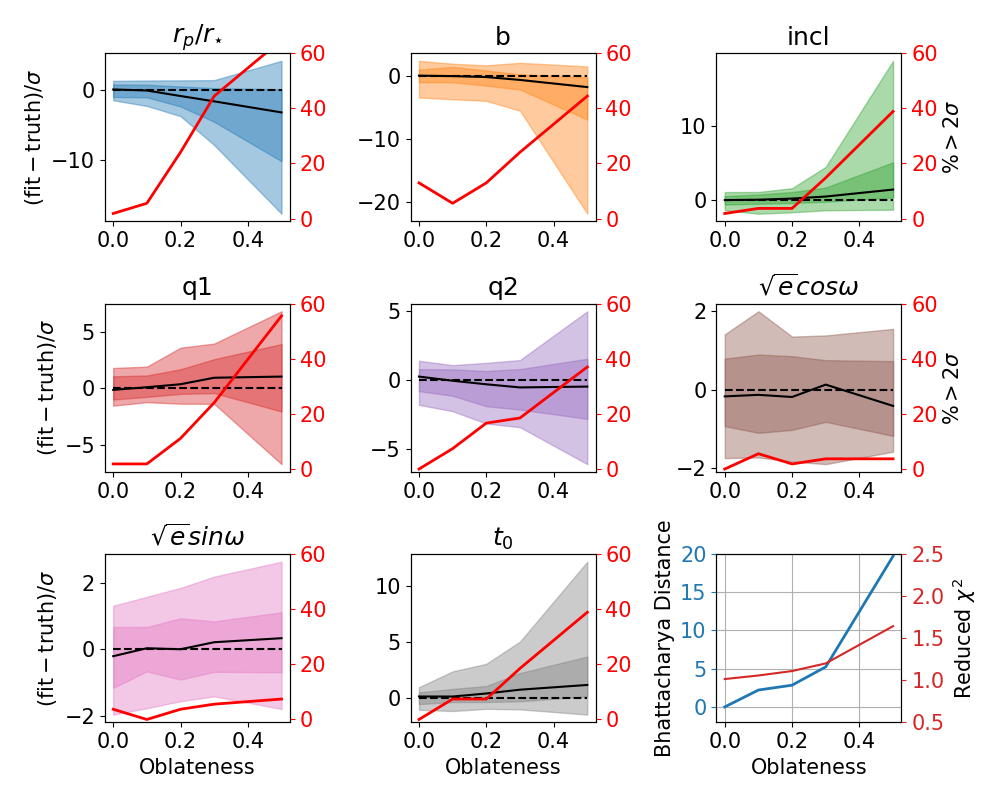}
	\caption{Results of fitting a spherical planet model to simulated data of an oblate planet, using orbital parameters for the 100 largest (in terms of planet to stellar radius) Kepler planets with periods greater than 10 days. The left vertical axis in each plot represents the number of standard deviations between the truth and fitted value for the planet to stellar radius, impact parameter, inclination, quadratic limb darkening parameters, eccentricity parameters, and transit center. Dark and light regions represent where ~65\% and ~95\% of the planets in the sample appear. The right vertical axis and solid red line show the percentage of planets for which the retrieved parameters deviate by $>$2 standard deviations. The bottom right panel shows the median values of the reduced $\chi^2$ in red, as well as the median Bhattacharyya distance in blue.}
	\label{fig:parameter deviations}
\end{figure*}

\subsection{Bias on retrieved parameters during a transit analysis}

Parameters of a model are often degenerate with one another, and given the low signal strength of oblateness it's possible that other transit parameters may be able to compensate and mask an oblateness signal at current SNR. Assuming a planet to be perfectly spherical could then introduce a bias when fitting for parameters such as inclination or the semi-major axis of the planets orbit. The goal of this section is to quantify such an effect through a series of simulation-retrieval studies. We note that in \cite{barnes:2003} a similar analysis was done to study the biases of assuming an oblate planet to be spherical. However in that case the effect was only studied for a single planet, HD209458b \citep{charbonneau:2000}, whereas our aim is to expand this analysis to a wider set of planets to understand the effect more globally.  We follow the methodology of \cite{dewit:2012} which looked at the degeneracies between the orbital properties of a planet and its potential to have non-uniform surface brightness, measured through minor variations during eclipse.

We first generate a set of lightcurves for a sample of oblate planets. The planet parameters we use come from the exoplanet archive list of confirmed transiting planets, as was done in the previous section. The parameter space of possible planet and orbital configurations is large, and so we restrict our study to known planets, which at the time of writing is a sample of greater than 4000 objects, and from these we take the 100 planets with the largest planet to stellar radius ratios (since the oblateness signal scales proportionally with radius). For those planets, we generate 1 minute cadence lightcurves with oblateness value from the range of 0, 0.1, 0.2, 0.3, and 0.5. We add Gaussian noise with a standard deviation of 50ppm for each data point. We then run a Monte-Carlo Markov-Chain (MCMC) fit to the synthetic data using the emcee package \citep{foreman-mackey:2012}, assuming a spherical planet model during the fit.

\subsection{Description of Fit}

We choose to fit for planet-to-stellar radius ratio, transit impact parameter, inclination angle, eccentricity, argument of periastron, as well as two parameters of a quadratic limb darkening model, using the q1 \& q2 parametrization of \cite{kipping:2013} and finally center of transit. We  hold the orbital period fixed. We note that some of these parameters, specifically the eccentricity, could be constrained by radial velocity observations which would be unaffected by planetary shape variations. Additionally, for the fits described in this section we hold the obliquity angle of the planet fixed at 45 degrees. We ran a similar analysis for a fixed obliquity of zero which we discuss further down.

Regarding the radius of the planet, we do not fit for it directly but rather scale it by the appropriate oblateness factor as previously mentioned. If we fit for the radius directly, we would see deviations on the order of 20\% or more as oblateness increased (in order to match the transit depth), however this would be somewhat misleading. While it is true that the equatorial radius would differ by such an amount, the surface area (and thus total occulted stellar area) would remain relatively unchanged. This means that a 20\% change in equatorial radius between a spherical and oblate planet does not represent the same magnitude of change between two spherical planets whose radii differ by 20\%.  By fitting for the scaled radius, any deviation we see is a more meaningful difference in quantities such as the total surface area or volume of the planet.

For a given planet we compare the posterior distributions of each fitted parameter to the known value that was used to generate the simulated data. We calculate the median value of the posterior distributions as well as the standard deviation. We then calculate by how many standard deviations the fitted parameter value is from the true parameter used when generating the data. If one assumes gaussian uncertainties for the fitted parameters, then for an appropriate model it would be expected that $\sim$68\% of the time the true parameter should be within one standard deviation, $\sim$95\% of the time within two standard deviations and $\sim$98\% of the time within three standard deviations. 

\subsection{Results of injection-retrieval tests}

The compiled outputs of our set of fits is shown in figures \ref{fig:corner plots} \& \ref{fig:parameter deviations}. As expected, for low oblateness values the fitted parameters cluster around the true value, with a spread of $\sim$ 2 standard deviations. This is the statistically expected result for a model which matches the data.

\subsubsection{Biases on radius, impact parameter, and inclination}

We observe significant biases with clear trends for the equatorial radius $r_p/{r_\star}$, impact parameter (b), and inclination (incl). For oblateness values greater than 0.2, we see that the first two of these parameters become biased towards smaller values, while inclination is biased towards larger values (i.e. the orbits tend towards the observer line of sight). We also see the distribution of fit parameters widen, which could be an indication that the errors on these parameters are being underestimated. 

For ($r_p/{r_\star}$) we find that 20\% of the fits deviate by $>$ 2$\sigma$ at f = 0.2, with this number increasing to $>$60\% as the oblateness increases. For the impact parameter we find that 10\% of the fits deviate by $>$ 2$\sigma$ increasing to 40\% at f = 0.5, and for inclination we find that from f = 0.2 to f = 0.5 the percentage of fits $>$2$\sigma$ rises from $\sim$5\% to 40\%.

A potential explanation for this is the effect of oblateness at the very beginning and end of transits. Compared to a spherical planet, and based on its orientation, an oblate planet will begin transiting slightly earlier and end transiting slightly later. This leads to a change in the overall duration of transit, which we see here is being compensated for by the three parameters which directly affect the length of the transit chord. 

In figure \ref{fig:corner plots} we note a significant correlation in particular between the impact parameter and inclination of the planet. We observe a similarly strong correlation between the inclination and radius ratio. 

%\textbf{
For the specific case of HD209458b, it was found in \cite{barnes:2007a} that when fitting a spherical planet model to a simulated lightcurve of an oblate planet the impact parameter tended towards a critical value of $b=0.707$ in order to account for the change in ingress/egress duration. We do not observe a similar trend for the larger sample of planets and parameters we have studied in this work. One important distinction between the two analyses is that we have included additional parameters in our fit, namely the inclination. The effect of this is that we have additional ways to account for the change in transit durations. Thus we find differing trends in the combination of impact parameter and inclination which can account for the variations between spherical and oblate lightcurve models.
%}
\subsubsection{Biases on limb darkening coefficients}

For both limb darkening parameters q1 and q2 there is no clear upwards or downwards trend, although we do see a widening of distributions (although with a smaller amplitude of deviations compared to the other parameters). If a change in transit duration is indeed the cause of any observed biases, this would make sense given that the limb darkening parameters are unable to alter the length of the transit chord and are less strongly biased. Despite a lack of a clear trend however, we still find that a significant number of fits deviate by more than 2 standard deviations, indicating that the limb darkening coefficients are still affected by the change in transit shape to some degree.

For q1 we find that 10\% of the fits deviate by $>$ 2$\sigma$ at f = 0.2, increasing to 60\% as the oblateness increases towards 0.5. For q2 we see a larger initial value, with $\sim$20\% of fits deviating by $>$ 2$\sigma$ at f = 0.2, increasing to $\sim$40\% at f = 0.5.

\subsubsection{Biases on transit center}

In the case of non-zero obliquity, we observe a significant shift in transit center for large oblateness values. As will be discussed in section \ref{sec:TTV}, this is an expected effect due to the asymmetry in ingress and egress for a tilted planet. As previously mentioned, we ran a similar analysis for a fixed obliquity of zero. In this case, the transit will always be perfectly symmetric about its center, and thus no shift in transit center is expected which is what was observed. We note here that the direction of the shift is always the same. This is due to the fact that the obliquity is the same for all planets, and we assign the impact parameter to positive values. Thus the asymmetry induced in the lightcurve will always be in the same direction with regards to ingress and egress. For f = 0.2 $\sim$ 10\% of fits deviate by $>$2$\sigma$, increasing to 40\% as f approaches 0.5.

\subsubsection{Biases on eccentricity}
For eccentricity and the argument of periastron $\omega$ (parameterized as $\sqrt{e}cos{\omega}$ and $\sqrt{e}sin{\omega}$) we find that the retrieved values are statistically consistent with the values used to generate the data (i.e. all samples are $<$3$\sigma$). We note however that this is likely to be a limitation of the data sets being analysed, which are single transit observations. Eccentricity can be more tightly constrained when additional information such as stellar density \citep{dawson:2012} and planetary eclipses are used \citep{winn:2007}. We thus would expect to obtain significant biases on these parameters from perturbed ingress/egress for the increased precision derived from joint fits of primary eclipses, secondary eclipses, and/or radial velocity measurements--similarly to the point made in Fig. 4 of \cite{dewit:2012}.

\subsubsection{Goodness of Fit Metrics}

In addition to having measured a clear bias for certain parameter when assuming a spherical planet model, we also check the goodness of fit statistic to asses if, given current precisions, there exists a range of oblateness leading to perturbations in the transit shape which cannot be compensated for by the orbital parameters. We show in the bottom right panel of figure \ref{fig:parameter deviations} the median value of the reduced $\chi^2$ across all planets in the sample for differing oblateness values. We note that this is measured only using data points during ingress and egress of transit, as the rest of the lightcurve is expected to be the same for both spherical and oblate planets. As shown, the value remains close to one, increasing only for values of oblateness approaching 0.5 (although still remaining below $\sim$1.5). In the same panel we also show the mean value of the Bhattacharyya distance ($D_B$), which is a metric for comparing the similarity of two distributions defined as 

\begin{equation}
	D_{B}(p,q) = -\mathrm{ln}\left(\Sigma_i \sqrt{p(\theta_i)q(\theta_i)}\right)    
\end{equation}

for two probability distribution p and q over a set of parameters $\theta$. 
The value shown for a given oblateness is a comparison between the multi-dimensional posterior distribution of that value and the posterior distribution for zero oblateness. We see a clear trend to higher values as the oblateness increases, indicating a statistically significant variation between the two distributions.  Thus we find that assuming planets to be spherical can statistically bias results in a way that does not set off the most common alarm for improper models.

%\begin{figure*}
%   \centering
%    \includegraphics[scale=0.7]{Transit_Shift.png}
%    \caption{The orange lines represent transit observations across several epochs. The blue line shows a single transit, and is used to indicate that the transits are altered in such a way as to appear to shift around their center, despite all models having the same center of transit time. This is due to the way in which the geometry of the planet interacts with the host star. (maybe add figure to right of this showing the geometry of the situation).}
%    \label{fig:transit shift}
%\end{figure*}

\section{The sensitivity of oblateness retrieval \& planet populations}
\label{sec:fitting}

\begin{figure*}[ht!]
	\centering
	\includegraphics[scale=0.9,width=.48\linewidth]{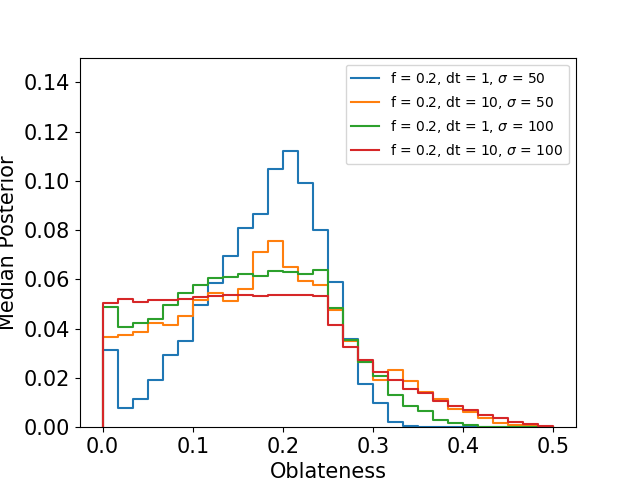}
	\includegraphics[scale=0.9,width=.48\linewidth]{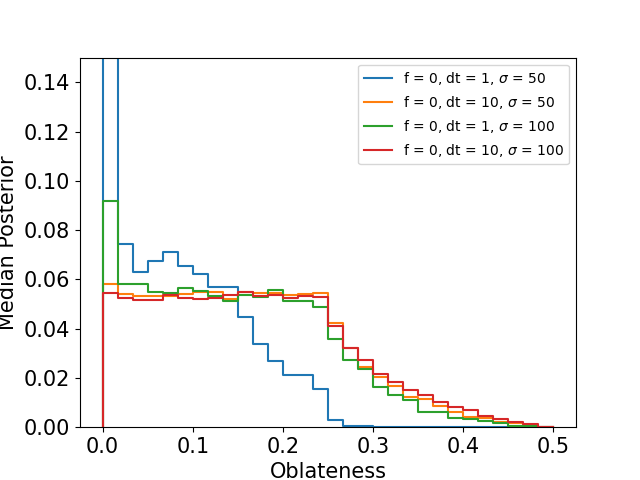}
	\caption{The results of MCMC retrievals to two populations of planets, using a model which accounts for oblateness and obliquity. On the left is a population where each planet has an oblateness value of 0.2, and the the right plot is a population of planets with zero oblateness. For each group, we test difference levels of observing cadence (dt in minutes) and Gaussian noise level ($\sigma$ in ppm)}
	\label{fig:synthetic posteriors}
\end{figure*}

Measuring the oblateness of an individual planet is difficult and requires very precise and well-sampled observations. The approach taken in this section is to study ensembles of planets in order to determine if planets can be separated into populations of oblate and non-oblate planets. For example, as given by equation \ref{eq:t_spin}, the tidal spin-down timescale varies strongly with orbital period (to the $\mr{4^{th}}$ power). Planets below a period of 15 days are expected to have very short spin-down timescales, implying they will be tidally locked to their host star and will have slowed down enough to have effectively zero oblateness. For planets with an orbital period above 15 days with much longer spin-down times, they are expected to have retained a high rate of rotation from their formation and thus have non-zero oblateness. In addition to searching for oblateness across a population, we also measure the sensitivity of oblateness fits to the quality of the data, which we characterise by SNR and time sampling of data points.

\subsection{Synthetic populations}

We generate 200 planets whose periods, radii, semi-major axis, and inclinations are drawn from the population of known transiting planets. For a given choice of observing cadence, oblateness factor, and noise level we simulate an oblate lightcurve as in previous sections. We do this process for two populations of planets with identical planet parameters, with one having zero oblateness and the other having an oblateness of 0.2 for all planets. 

Once we have these lightcurves, we run an MCMC retrieval on them using the Allesfitter package \citep{allesfitter-paper, allesfitter-code}. We include in Allesfitter our own oblateness model, which allows us to fit for the oblateness parameters f (oblateness) and $\theta$ (the obliquity of the planet).

We show in figure \ref{fig:synthetic posteriors} the results of this analysis, in particular the posterior distributions retrieved for the oblateness. We note that in our analysis we reparametrize f and $\theta$ as  $\sqrt{f}\mathrm{cos}\theta$ and $\sqrt{f}\mathrm{sin}\theta$, similar to the parametrization often used when fitting for eccentricities \citep{vaneylen:2015}. These reparametrized versions of the shape parameters do not have hard boundaries at 0 and instead vary from -1 to 1, which removes the bias of forcing oblateness to be positive.

We first note that in the case of zero oblateness (the right panel), the general trend is a flat posterior below a certain value, in this case around $f = 0.25$, followed by a sharp decrease beyond that. The model is able to confidently rule out high oblateness values, which in turn can be used to rule out large rotational periods of planets. Below the cutoff point the model is less sensitive to oblateness. We see that for the most constraining set of parameters, namely a short observing cadence of dt = 1 minute and a small noise level of 50 ppm, the model is able to exclude smaller oblateness values more confidently. 

In the left panel of figure \ref{fig:synthetic posteriors} we see two cases emerge, depending on the noise properties and time sampling. For the shortest time sampling and lowest noise level, the model is able to retrieve the population oblateness level of 0.2. As the noise level increase along the with the time sampling, the posterior gradually shifts to that of a zero oblateness population. Thus we see that between 50-100 ppm, and between 1min - 10 min observing cadence there is a cutoff point where oblateness shifts from being confidently detectable to completely undetectable.

\subsection{Sensitivity of oblateness fits to data quality}
\label{sec:sensitivity}

In order to investigate the shift in retrieval seen, we sample a grid of data quality, which we characterise by two metrics. One metric is the signal-to-noise ratio, which in our case we take to be the transit depth divided by the amplitude of Gaussian noise. The other value we use to quantify the data quality is the number of data points observed during ingress, which is calculated as the duration of transit divided by the observing cadence. For each data point sampled, we run an MCMC retrieval as described above, and calculate the oblateness value which encompasses 95$\%$ of the posterior distribution (which is often asymmetric due to the boundary at f = 0).

\begin{figure}[h!]
	\centering
	\includegraphics[scale=0.35]{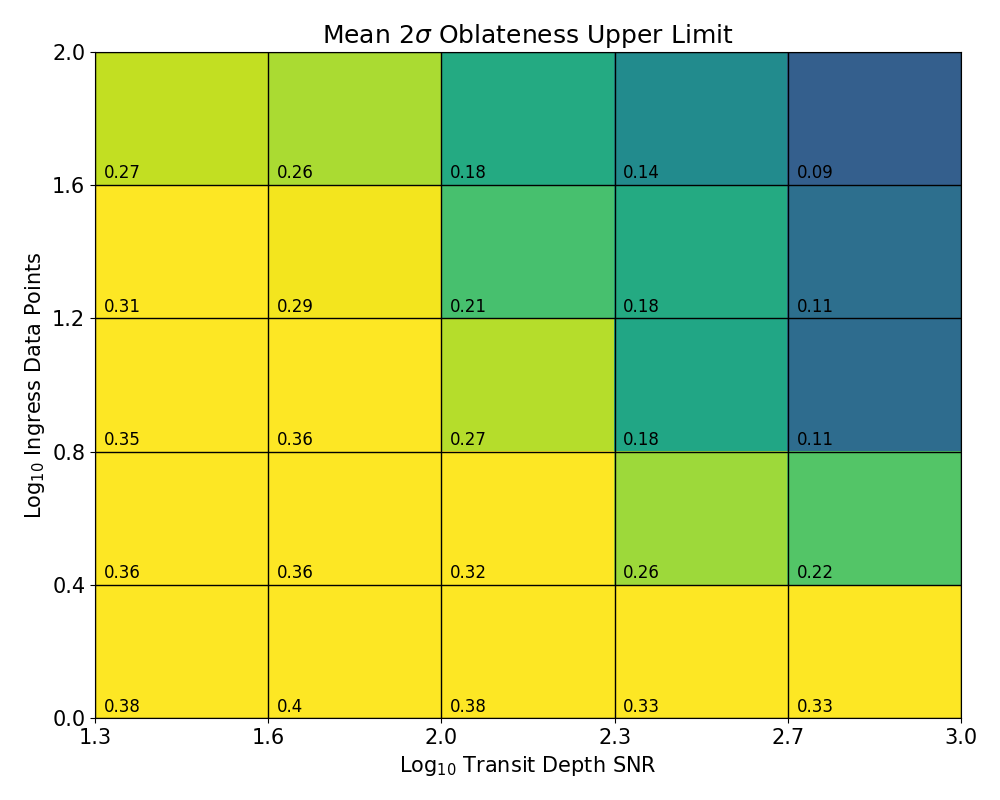}
	\caption{The median 2$\sigma$ oblateness upper limit values across a sample of planets retrieved when fitting an oblate planet model to synthetic data sets with zero oblateness (values shown in grid cells). The x axis indicates the transit depth divided by the amplitude of Gaussian noise. The y axis shows the duration of transit ingress divided by the observing cadence.}
	\label{fig:sensitivy map}
\end{figure}

The results of this are shown in figure \ref{fig:sensitivy map}. In this case the data being fit has an oblateness value of zero, i.e. spherical planets. Thus we are demonstrating the ability of the model to confidently rule out non-zero oblateness. We see that for low SNR values and a low number of transit data points, the model is unable to rule out values of oblateness below 0.3 at the one standard deviation level, and f = 0.4 at the 2$\sigma$ level. It is only for very high SNR values $>$ several hundred and more than at least 10 transit points that the model is even able to rule out oblateness values of 0.2 or greater. We note that the oblateness value within each cell varies to a certain degree amongst the different planets. This is due to the fact that properties such as the impact parameter also play a role in measuring oblateness, however here we focus on the bulk results marginalized across such parameters.

The significance of this is that the majority of transiting planet observations are completely degenerate between a spherical planet and a planet having an oblateness of $\lesssim$ 0.25. In the context of section \ref{sec:density} for example, this implies that analysis which rely on relative density errors below $\sim$ 15$\%$ are over-confident, in the case of planets which have the potential to be oblate (i.e. planets which are unlikely to be tidally locked).

\subsection{Oblateness fit of long and short period Kepler planets}

\begin{figure}
	\centering
	\includegraphics[scale=0.45]{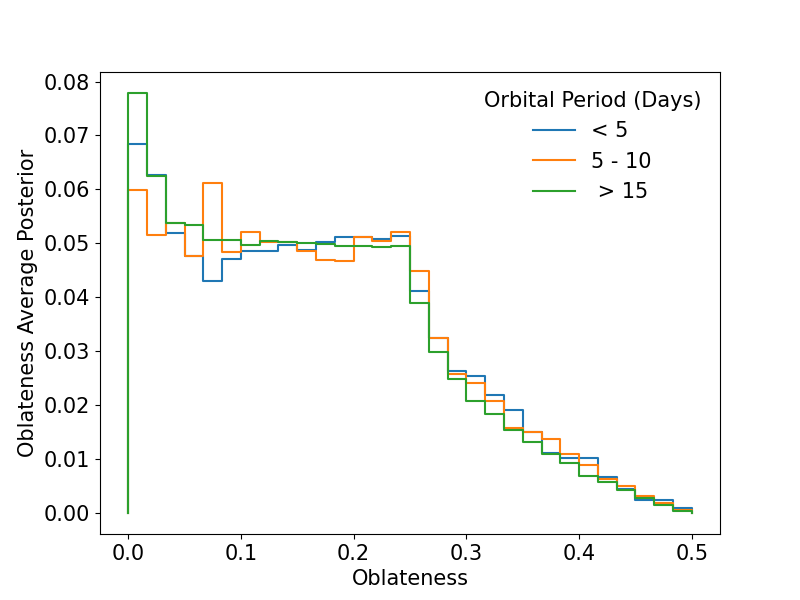}
	\caption{Oblateness posteriors when fitting confirmed Kepler planets, separated by orbital periods.}
	\label{fig: population posteriors}
\end{figure}

We now turn to the potential oblateness of planets discovered by the Kepler survey. The reason for picking this as our sample is the large number of planets it contains, the fact that many of them were observed at a short (1 minute) cadence, the duration of the survey which provides a large number of transits for most planets, and finally the SNR it provides for most planets, which far surpasses ground based and many spaced-based transit observations.

In contrast to previous studies which have searched for oblateness in individual planets, our goal is to see if it is possible to highlight populations of planets which globally either do or do not exhibit oblateness. Given that planetary rotation is the root of the signal, we propose to separate planets into fast and slow rotators, which would then translate into populations of high and low oblateness planets.

We download the short cadence lightcurves of all planets with radii larger than a cutoff value of 6 $R_e$, to exclude rocky planets with low tidal dissipation factors (and thus extremely short circularization timescales). We do this using the lightkurve package \citep{lightkurve}, and extract the PDCSAP lightcurves. We then run them through a suite of post-processing and normalizing steps. We first remove points which have bad data quality labels. We next identify all transits of a system, including those with multiple planets. We chose not to exclude these systems, although we do remove observations where two transits come within a multiple of 3 transit durations of one another. We then apply a Savitzky-Golay filter to remove long term variations in the lightcurve. We mask the known transits while doing this in order to not have them bias the detrending. We next correct for systems which exhibit large TTVs, which in some cases by on the order of tens of minutes. We do by this running an MCMC fit for each individual transit, holding fixed all known transit parameters and allowing only the transit center and depth to vary. 

Once we have measured transit centers for all planets, we repeat the previous detrending steps, using the shifted transit centers to accurately mask out the transits when fitting for low frequency lightcurve trends. Once this is done we phase fold the data into a single lightcurve, which is what is finally used in the fitting routine. We run an MCMC fit to the data utilizing the Allesfitter package as we did in the previous sections.

We show in figure \ref{fig: population posteriors} the stacked posterior distributions for planets with periods either above or below 15 days. We find similar distributions as in the previous section for the cases in which the fit was unable to identify significant oblateness. The fact that both distributions look similar indicate that the sample of Kepler planets does not exhibit an oblateness bias or grouping based on their period. We analyzed different groups of periods, cutting below 10 days or above 20, and similarly found no biases or differences between any groups. 

We also see the same cutoff at oblateness values above $\sim$0.25, again as in the previous section. This indicates that the MCMC disfavors such high levels of oblateness. High oblateness values in general are unlikely, due to the projection effects described in section \ref{sec:model} and Appendix A. In section \ref{sec:known planet ob level} we showed the oblateness signal induced among known planets, and show that very few planets produce a signal beyond 100ppm. However for many planets in the Kepler survey the noise level is on the order of 100s of ppm. Thus while we are able to exclude high values of oblateness in general, we find that the sensitivity factors described in the previous subsection limit the ability to detect or rule out more moderate values of oblateness.

We do note a slight increase in the posterior for oblateness values $<$ 0.05 for planets with orbital period $>$ 15 days. This is again most likely due to increased sensitivity discussed in the previous section, given that planets with larger orbital periods will have increased sampling during ingress / egress and also tend to be larger. Thus we do not attribute this as indicating that larger period planets have lower oblateness.

\section{discussion}
So far we have studied the effects on standard transit analysis when allowing for an exoplanet to be oblate instead of spherical. If not properly taken into consideration, ignoring the potential of a planet to be oblate could introduce additional biases into more sophisticated exoplanet analysis. This highlights a main point of this work, which is that as the quality of data continues to improve, features such as oblateness which were once undetectable may begin to bias and effect a wider range of conclusions one may make about explanatory systems. In this section we consider the measurement of bulk densities of planets, as well as transit timing variations and the ways in which oblateness may impact any conclusions drawn. We also describe a metric for quantifying which planets are more or less likely to exhibit oblateness, based not only signal strength but also on their likelihood to be oblate in the first place.

\label{sec:disc}
\subsection{The relative uncertainty of bulk densities}
\label{sec:density}
When discussing radii in section \ref{sec:biases}, care was taken when comparing a spherical planet described by just one radius to an elliptical planet with two radii. However for a quantity such as the density of the planet, there is less of a distinction since the bulk density is typically the quantity of interest. We will illustrate that any analysis which utilizes precise density measurements may be subject to biases due to the shape of the planet.

Consider a planet which has an oblateness f, which induces oblateness at a level that is too low to be detectable or even to induce the biases previously mentioned. The depth - radius relation for an oblate planet is given by $\delta = r_{eq}r_{pol} = r_{eq}^2(1-f)$. This would translate to a measured planet volume of

\begin{equation}
	Vol_{ob} = \frac{4\pi}{3}r_{eq}^2r_{pol} = \frac{4\pi}{3}r_{eq}^3(1-f) = \frac{4\pi}{3}\frac{\delta^{3/2}}{(1-f)^{1/2}}
\end{equation}

Assuming mass measured by radial velocities to be independent of shape, we can compare the inferred volumes of a spherical and oblate planet, and thus their densities, as

\begin{equation}
	\label{eq:density variation}
	\frac{\rho_{circ}}{\rho_{ob}} = \frac{4\pi}{3}\frac{\delta^{3/2}}{(1-f)^{1/2}} \frac{3}{4\pi}\frac{1}{\delta^{3/2}} = \frac{1}{(1-f)^{1/2}}
\end{equation}

For an oblateness of even just $f=0.1$, this would translate to a true density that is 5\% smaller than the density inferred for a spherical planet. An oblateness of $f=0.2$ would mean a discrepancy of almost 11\%. 

\begin{figure}
	\centering
	\includegraphics[scale=0.46]{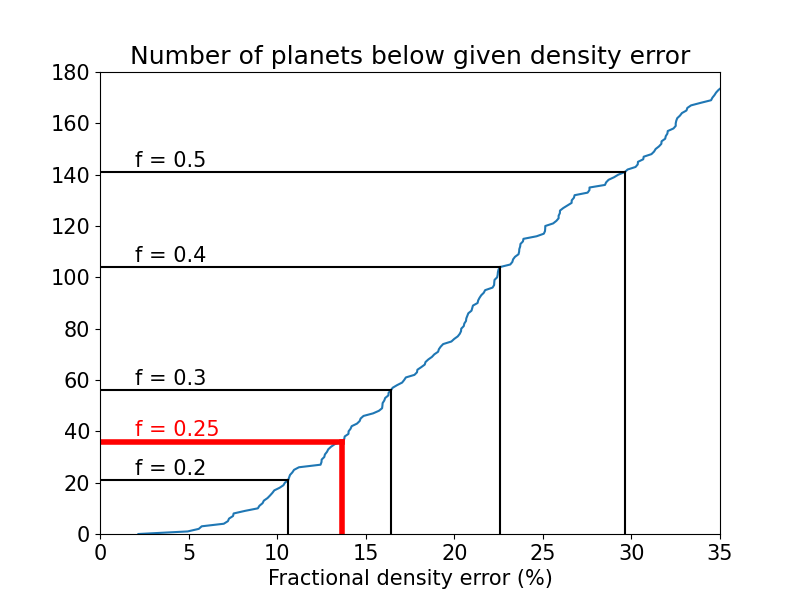}
	\caption{The cumulative number of planets with a fractional density error below a certain value (blue line). The black lines show the percentage by which the density would change if a previously assumed spherical planet were to be oblate to a certain degree, calculated using equation \ref{eq:density variation}. The red line indicating a value of f = 0.25 is in reference the limits of current observations discussed in section \ref{sec:sensitivity}.}
	\label{fig:density errors}
\end{figure}

In figure \ref{fig:density errors} we compare this oblateness induced density variation to the relative uncertainty of the density of known planets. We take all planets from the exoplanet archive which have a reported value of both physical radius and mass, as well as reported uncertainty values. We calculate their relative density errors, and find for an oblateness of 0.2, up to 20 planets would have their density differ by more than their relative uncertainty. For an oblateness of 0.3, that number rises to almost 60. It was demonstrated in section \ref{sec:sensitivity} that at low data sampling and/or transit SNR it is impossible to rule out oblateness values less than 0.3 at the 2$\sigma$ level. This in turn imposes an accuracy wall on relative uncertainty of a planet's bulk density at the level of 10$\sim$15\%.

\subsection{TTV-like Signal induced by Oblateness}
\label{sec:TTV}
\begin{figure*}[ht!]
	\centering
	\begin{subfigure}[t]{0.48\textwidth}
		\centering
		\includegraphics[width=\textwidth]{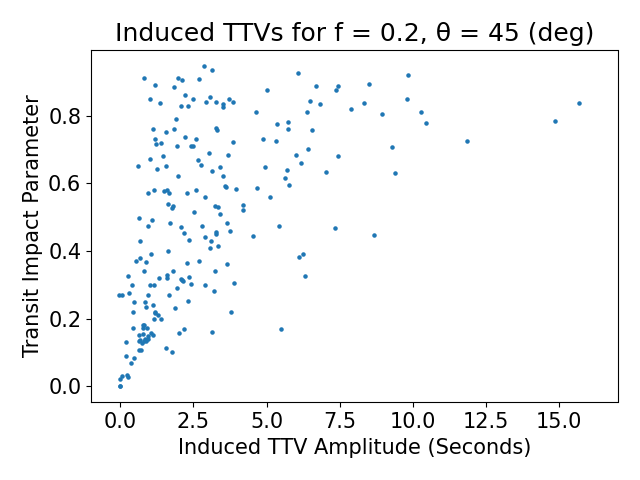}
		\caption{The distribution of retrieved TTV amplitudes induced by asymmetry stemming from oblate planets with f = 0.2 and $\theta = \pi/4$. The y axis shows the impact parameter of the transiting planets, which we expect to be correlated with the asymmetry signal.}
		\label{fig: ttv amplitude vs impact par}
	\end{subfigure}
	\hfill
	\begin{subfigure}[t]{0.5\textwidth}
		\centering
		\includegraphics[width=\textwidth]{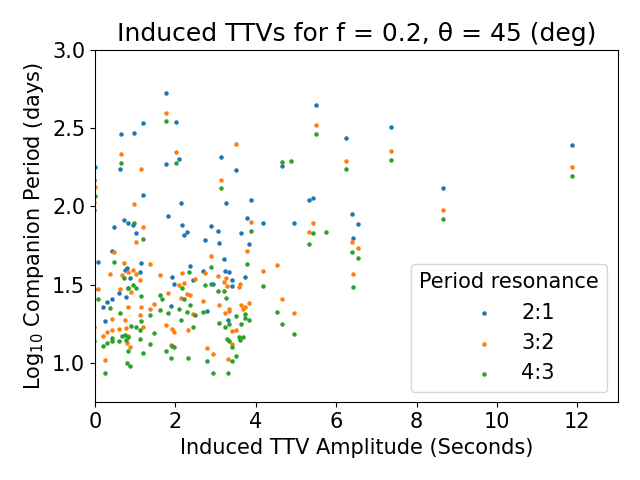}
		\caption{The orbital periods of a companion that would be needed to induce the TTV signals seen, based off the precession period and possible orbital resonance.}
		%\caption{The distribution of TTV-like amplitudes retrieved as a function of the obliquity angle of the planet. For low amplitude planets there is not much of an effect, but we see a noticeable decrease for larger amplitude planets as the obliquity angle deviate from 45 degrees relative to the orbital plane of the planet.}
		%\label{fig:effect of obliquity on retrieved amp}
		\label{fig:companion period}
	\end{subfigure}
	\caption{The oblateness induced TTV effect}
	\label{fig:}
\end{figure*}

Thus far we have considered the oblateness of a planet to be constant, however it is possible that a planet with a non-zero obliquity angle will precess over time as it orbits its host star and exchanges angular momentum. For uniform precession in a fixed orbit, the precession period is given by \citep{ward:1975}

\begin{equation}
	\label{eq:p_prec}
	P_{prec} = 13.3\ \mr{years}\times \left( \frac{P_{orb}}{15\ \mr{days}}\right)^2\left(\frac{10 \mr{hr}}{P_{\mr{rot}}} \right)\left( \frac{\lambda/J_2}{13.5}\right)\frac{1}{\mr{cos}\theta}
\end{equation}

where the value of 13.5 comes from the estimate of $\lambda / J_2$ for Saturn \citep{ward:2004}. In this equation $\theta$ is the obliquity angle of the planet, which we recall is the angle between the polar axis of the planet and its orbital plane. As mentioned in section \S \ref{sec:model}, as the orientation of an oblate planet changes in space, an observer will see different elliptical projections on the sky plane. This will lead to a time varying projected oblateness and obliquity, which will in turn lead to a time varying transit shape. 

The focus of \cite{biersteker:2017} was to try and detect oblateness due to do variations in the depth of a planets transit, caused by a time varying projected surface area (including a change in both the projected oblateness as well as equatorial radius). A secondary effect which was first mentioned in \cite{carter:2010b} is that the duration of the transit may change as well, which could mimic the appearance of a transit timing variation (TTV) signal.

If an oblate planet has a non-zero impact parameter and obliquity, then it will exhibit asymmetry in ingress and egress (as in figure \ref{fig:oblateness diagram}). Thus when attempting to fit a spherical transit model to the lightcurve of an oblate planet, the result could be a perceived shift in the center of the transit. This effect is typically attributed to another planet in the same system gravitationally perturbing the observed planets orbit \citep{agol:2018}. Thus attempting to fit a spherical planet model to a precessing oblate planet could potentially lead one to conclude the existence of an additional planet causing timing variations.

We perform a set of injection retrieval tests, attempting to measure a shift in the center of transit. We simulate a transit lightcurve of an oblate planet across multiple epochs while altering the obliquity over time according to its precession period. As the true obliquity of the planet changes over time, we also update the projected obliquity and projected oblateness. We must additionally account for the change in projected equatorial radius of the planet, which we do using equation 4 of \cite{carter:2010b}. The formula provides the relative transit depth as a function of oblateness and inclination parameters.

We use the same sample of planets as in previous sections in order to determine the degree to which the current population of planets may be affected. We note that the precession period given in equation \ref{eq:p_prec} is only an estimate, and further effects such as interactions with other planets in the same system could disrupt or alter the precession in someway \citep{saillenfest:2019}. We hold the parameter ${\lambda/J_2}$ fixed to the value of 13.5 for Saturn, and note that the precession period scales linearly with it. We sample the obliquity of the planet from three value spaced out between 0 and PI/2. In the case of of either zero or maximal (PI/2) obliquity, there would be no ingress/egress asymmetry and thus no induced TTV signal. For these values we find that most precession periods range from 10 - 100 years, although there are some planets with potential periods of only a couple years or less. For most planets such an effect will thus occur on too long of a timescale to be observable. As the field of exoplanet enters its $4^{th}$ decade however, it becomes more and more feasible to observe such a long baseline transit shift and falsely attribute it to a long period companion.

Many of the trials return a constant transit center as the best fit, indicating no detectable variation. For a significant number of cases however the best fit is a clear sinusoidal variation in the transit center, as one would expect from TTVs. For these cases we fit a sinusoid to the time varying signal and retrieve the amplitude of the variation. In figure \ref{fig: ttv amplitude vs impact par} we show the distribution of these amplitudes, in units of seconds, as a function of the impact parameter of the transit. We note a correlation with the impact parameter, which is extremely tight for values less than ~0.15, and then widens above that. We find that for an underlying oblateness of 0.2 and an obliquity of 45 degrees, the scale of the deviations is on the order of several seconds, with the peak value being upwards of 15 seconds (and thus 30 seconds peak to peak). While small, transit deviations have been measured at the level of seconds, for example in Wasp-12 b \citep{maciejewski:2016}.

In figure \ref{fig:companion period} we show the possible orbital periods of the companion planets that would be required to induce the observed TTV signal. This is calculated using equations 6 and 7 of \cite{lithwick:2012}, where we substitute the precession period of the oblate planet for the super-period of the TTV signal. We calculate a companion period under the assumption of a 2:1, 3:2 and 4:3 period resonance. The periods we find span the range of 10 - 1000 days.

As mentioned we repeated the above analysis while varying the obliquity angle of the planet. This was done to ensure the results are consistent with the expectation that for obliquity angles further from 45 degrees the induced amplitudes should be smaller due to the reduced asymmetry between ingress and ingress. This is indeed what was found, and in particular for an obliquity of zero/90 degrees (in which the entire transit is perfectly. symmetry about its midpoint) there were no statistically significant transit center variations.

\subsection{A metric to identify prime oblateness candidates}
\label{sec:metric}

In this section we define an oblateness observability metric to determine which planets are the most ideal for a targeted study of oblateness. It is built in such a way to favor planets which would not only have a large oblateness signal, but which are also likely to be oblate in the first place. The key factor we use to determine if a planet is likely to be oblate or not is the age of its host star. 
%\textbf{
As a planets rotation slows down over time due to tidal de-spinning, its oblateness similarly decreases. The timescale for a planet to spin down and become tidally locked can be estimated by:
%}

\begin{equation}
	\label{eq:t_spin}
	\begin{split}
		\tau_{spin} = 1.22\ \mathrm{Gyr} \times \left(\frac{M_p}{M_{J}}\right)\left(\frac{Q_p}{10^{6.5}}\right)\left(\frac{\lambda}{0.25}\right) \\
		\times \left(\frac{P_{\mathrm{orb}}}{15\ \mathrm{days}}\right)^4\left(\frac{R_J}{R_eq} \right)^3\left(\frac{10 hr}{P_{\mathrm{rot,i}}} - \frac{10 hr}{P_{\mathrm{rot}}}\right)
	\end{split}
\end{equation}

which is derived by integrating the spin down rate of a planet as described in \cite{goldreich:1966}. The quantities $Q_p$ and $\lambda$ are the tidal dissipation factor and normalized moment of inertia $\lambda = I / (M_p R_p^2)$. If the mass of the planet is known, one can additionally directly calculated the rotational period required for various levels of oblateness.

%\textbf{
Given the uncertainty and difficulty in calculating Q, we conservatively set it to $10^{6.5}$, which is above the estimate for the solar system values of Saturn and Jupiter, which are typically on the order of $10^4$ and $10^6$ respectively \citep{storch:2013}. By over-estimating Q, we in turn will likely be over-estimating the tidal spin-down timescale. Thus, we can be confident that systems which are older still than this estimate will have already spun-down and are unlikely to be oblate. For the parameter $\lambda$ we adopt the commonly used value of 0.25 for gas giants. As a comparison, the Darwin-Radau equation\footnote{$\frac{J_2}{f} = -3/10+5/2\lambda-15/8\lambda^2$} gives a value of $\lambda = 0.22$ for Saturn, and $\lambda = 0.27$ is calculated for Jupiter from Juno gravity measurements \citep{ni:2018}.
%}

%Assuming conservative values for Q and $\lambda$, we are able to estimate an upper limit for this timescale based only on the period, equatorial radius and, if it is known, the mass of the planet. Thus for cases where even such an upper limit for the circularization timescale is below the stellar age, oblateness can be safely ruled out. 
%\textbf{
Conversely, planets whose associated star has an age that is well below 1 Gyr can be assumed to not have had enough time to spin down and may still have a high rotation gained during formation.
%}
We incorporate this information using a sigmoid function, defined as $\sigma(x) = 1 / (1+e^{-x})$ which is a smoothed out step function. We scale the step width by the uncertainty of the stellar age, to allow for flexibility for spin down estimates near the stellar age.

We multiply the factor for the timescale by both the RMS variation induced by oblateness (as described in section \ref{sec:known planet ob level}), and additionally by a term which is proportional to the expected SNR of a detection, for a given waveband. This takes the form

\begin{equation}
	\begin{split}
		Metric = RMS_{ob} \times 10^{-M/5} \times \sigma((Age_{\star}-\tau_{spin}^{upper})/\sigma_{Age_{\star}})
	\end{split}
\end{equation}

Where M is the magnitude of the host star in a given waveband, and the exponential term signifies the square-root SNR estimate of a given number of photons. The $RMS_{ob}$ term is the same ingress/egress deviation measured in the section \ref{sec:known planet ob level}.

\begin{table}[]
    \footnotesize
	\centering
	\begin{tabular}{ccccc}
		\hline
		Planet &
		\begin{tabular}[c]{@{}c@{}}Period\\ (days)\end{tabular} &
		\begin{tabular}[c]{@{}c@{}}Radius\\ (Rj)\end{tabular} &
		K-Mag &
		\begin{tabular}[c]{@{}c@{}}Ob. Amp\\ (ppm)\end{tabular} 
		
		\\ \hline
		TOI-1899 b   & 29.02 & 1.15 & 10.51 & 110.45 \\
		TOI-1278 b   & 14.48 & 1.09 & 9.74  & 57.67  \\
		Kepler-699 b & 27.81 & 1.46 & 13.67 & 104.8  \\
		CoRoT-10 b   & 13.24 & 0.96 & 11.78 & 66.27  \\
		TOI-837 b    & 8.32  & 0.77 & 8.93  & 29.19  \\\hline
	\end{tabular}
	\caption{Top ranked planets according to oblateness metric}
	\label{tab:ranked-planets}
\end{table}

\section{Conclusions}

In this work we have analyzed the effect of planetary oblateness on the analysis of transit lightcurves, primarily through the use of injection-retrieval studies of spherical / oblate planets using spherical / oblate transit lightcurve models. %\textbf{
We first estimated the level of variation between the lightcurves of a spherical and oblate planet which otherwise have identical orbital and planetary parameters. We find that for most planets with periods greater than 10 days, this variation is at the 10-100ppm level.
%}
This result is obtained across a sample of planets which range in size, orbital period, impact parameter, and semi-major axis, which all affect the structure of an oblate planets lightcurve.

%We first determined that in the range of 10-100 ppm, as many as 100 of the currently know transiting planets with orbital periods greater than 10 days could have detectable oblateness levels. This result is obtained across a sample of planets which range in size, orbital period, impact parameter, and semi-major axis, which all affect the structure of an oblate planets lightcurve.

Given that this is a level at which instruments such as \textit{Kepler} are capable of reaching, we then studied effects which may obscure oblateness signals, notably compensation by orbital parameters \citep{barnes:2003,dewit:2012}. In addition to SNR, we also analyzed the effects of time sampling on the ability to robustly measure or rule out oblateness. What is found is that ruling out Saturn like oblateness (f = 0.1) at the 2$\sigma$ confidence level requires a sampling of $>$ 10 data points during ingress/egress and a transit depth which is $\geq$ 100 times the level of uncorrelated Gaussian noise. Combining both of these effects implies that for many planets there will be a sensitivity wall which limits the ability of a retrieval method such as MCMC to detect oblateness. When analyzing a sample of short cadence Kepler observations, we find this limiting factor to be in the range of f = 0.15 - 0.25. A consequence of this result is a limit on the accuracy of the bulk density which can be measured for planets which are potentially oblate. We find that relative density uncertainties below 10$\%$ for planets with periods $>$ than 10 days are likely to be overestimated, given the difficulty in ruling out oblateness levels below f = 0.2. If we extend this statement to planets which have been distorted through other processes such as tidal bulging, then the effect may be prevalent for planets with shorter periods as well.

In addition to density, we also find that attempting to fit spherical transit models to data which come from oblate planets produces fits which are both statistically consistent with regards to the reduced $\chi^2$, but which also deviate significantly from the values used to generate the data. For an oblateness approaching 0.5, we do find that the reduced $\chi^2$ begins to approach 2, indicating that the model is insufficient for the data. However in the range of f = 0.1 - 0.3, we find that it remains close to 1, while the Bhattacharyya distance becomes $\gtrsim$ 3 indicating deviation from the true parameter distributions. 

The planet radius, inclination, impact parameter, and transit center all demonstrate biasing trends as oblateness increases, while the two parameters of a quadratic limb darkening law deviate symmetrically about the true value. The eccentricity and argument of perihelion however remain unbiased and consistent with the truth values, even for large values of oblateness. This is expected given that we do not consider additional information such as secondary eclipses or radial velocity observations, which would provide a much more accurate measurement of eccentricity (in which case we would expect biases to appear). Additionally, time variation of oblateness due to precession is capable of mimicking a transit timing variation signal. We note however that this variation would be over significant timescales ($>$ 1 year) and only at the level of 10's of seconds.

When considering the biases and degeneracies we have found for oblateness, we note that there is the potential to extend this to other sources of shape variation. The models of \cite{barnes_j:2009} describe the effect of winds on an exoplanets atmosphere, which would distort its outer shell into a myriad of potential shapes which may each bias a lightcurve fit in different ways.

%On the other hand, we have also shown that for a targeted search of oblateness, going after the best of targets, there are several potential markers and signs of non-spherical planet which may likely be present and cannot be avoided. Furthermore, when attempting to study a particularly subtle feature of a planet, such as the map of its surface, whose signal would be highly degenerate and convolved with shape variations, at levels below 100ppm it would be a critical mistake to ignore the potential of a planet to deviate from a perfect sphere.

Furthermore, atmospheric studies of exoplanet surfaces, such as those which leverage the eclipsing of a planet by its host star, could benefit significantly by having the properties of the planets shape be pinned down precisely through an analysis of its transit, where the signal is often much stronger. Alternatively, if a planet which is potentially oblate is assumed to be spherical, this may further bias the results of any higher order analysis which is based off an understanding of the orbital and physical properties of a planet derived from transit observations.

\section{Acknowledgements}
DB
acknowledges support from an NSERC PGS-D scholarship as well as an FRQNT Doctoral Research Scholarship.

\bibliography{ms}

\pagebreak
%\onecolumngrid
%\appendix
		
	%\section*{Appendix A}
	%\label{appendix:proj}
	%\label{appendix:a}
	\section*{appendix: The Effect Of Projection}
	When observing an oblate planet what is seen is a two-dimensional projection, which is itself still an ellipse which can be characterised by a projected oblateness and projected obliquity. We consider the amount by which the true oblateness is altered if we assume a planet to have any uniform orientation in space. Note that it is only possible for projection to decrease the oblateness factor. When considering re-orientations of the planet, which has two long major axes A and a short major axis $B = A(1-f)$, the radii of the cross section of any rotation must be larger than B and smaller than A. Namely the projected short axis b must obey $B_{\mr{proj}} \geq B$ and the project long axis $A_{\mr{proj}} \leq A$. Recalling the definition of oblateness we find
		
	\begin{equation}
		f_{\mr{proj}} = 1-\frac{B_{\mr{proj}}}{A_{\mr{proj}}} \leq 1-\frac{B_{\mr{proj}}}{A} \leq 1-\frac{B}{A} \leq f
	\end{equation}
		
\begin{figure*}[ht
		]
		\centering
		\includegraphics[scale=0.5]{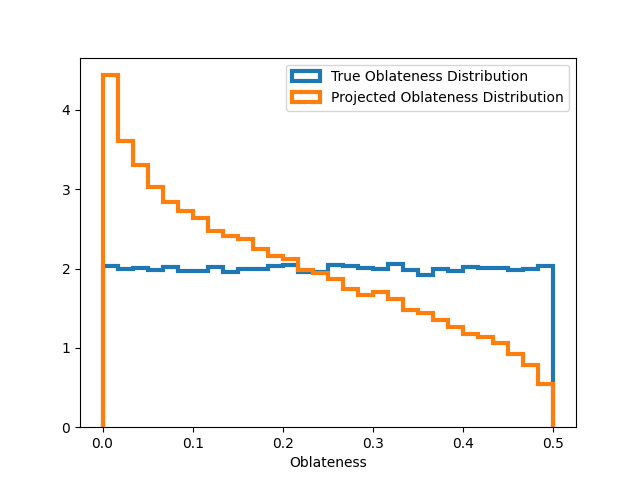}
		\caption{The effect of projecting a three-dimensional ellipsoid onto the sky plane, when its rotational axis is uniformly distributed across the unit sphere. As expected there is general shift towards lower oblateness values. This implies an additional level of difficulty when attempting to measure oblateness, since even if a planet were to have a high oblateness it would present itself as closer to spherical. Utilising a more informed model for the orientation of planets would provide a more accurate description of the transformation, however this would still be plagued by uncertainty about the true underlying oblateness distribution.}
		\label{fig:distribution projection}
	\end{figure*}

	When considering random rotations, in figure \ref{fig:distribution projection} we show the average transformation of a uniform distribution of oblateness values. As expected we see a noticeable shift towards shorter oblateness values. In this case we have chosen the two orientation angles of the planet to be uniformly distributed across the unit sphere. Analysing a large number of random projections, we find that any given oblateness value is on average decreased by a factor of about 20\%. 
		
\begin{figure*}
		\centering
		\includegraphics[scale=0.5]{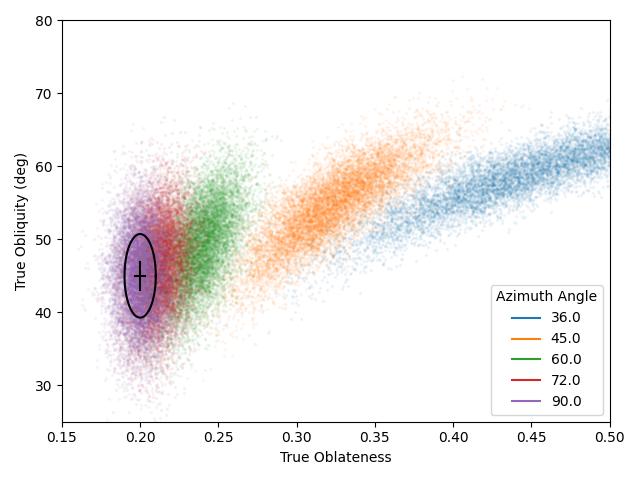}
		\caption{The solid black cross and oval represent the 1$\sigma$ measurements of a potential planet's (projected) oblateness. The coloured patches represent various distributions for the true (i.e. un-projected) values of oblateness and obliquity which would match the projected observation, for different choices of the azimuthal angle.}
		\label{fig:inverting distributions}
	\end{figure*}

	We may also ask, given a measurement from say an MCMC fit to a planet in which one retrieves a best fit value along with an uncertainty for the (projected) oblateness and obliquity, what possible underlying distribution of the true shape properties would match the observed values. For a choice of the azimuthal angle, there is a unique mapping between the projected and un-projected values, which can be obtained using equations \ref{eq:f_proj},\ref{eq:theta_proj},\ref{eq:theta_prime}. The results of this are shown in figure \ref{fig:inverting distributions}. For an azimuthal angle of 90 degrees we find that the two distributions are the same. In this case the rotational axis of the planet is tipped towards the observer, thus the two semi-major axes seen are always the true long and short axis (implying $f_{\mr{proj}} = f_{\mr{true}}$) and the obliquity angle is similarly observed to be the same as the un-projected value.

\end{document}